# PREDICTION OF REMAINING LIFE OF POWER TRANSFORMERS BASED ON LEFT TRUNCATED AND RIGHT CENSORED LIFETIME DATA

By Yili Hong, William Q. Meeker and James D. McCalley

*Iowa State University*

Prediction of the remaining life of high-voltage power transformers is an important issue for energy companies because of the need for planning maintenance and capital expenditures. Lifetime data for such transformers are complicated because transformer lifetimes can extend over many decades and transformer designs and manufacturing practices have evolved. We were asked to develop statistically-based predictions for the lifetimes of an energy company's fleet of high-voltage transmission and distribution transformers. The company's data records begin in 1980, providing information on installation and failure dates of transformers. Although the dataset contains many units that were installed before 1980, there is no information about units that were installed and failed before 1980. Thus, the data are left truncated and right censored. We use a parametric lifetime model to describe the lifetime distribution of individual transformers. We develop a statistical procedure, based on age-adjusted life distributions, for computing a prediction interval for remaining life for individual transformers now in service. We then extend these ideas to provide predictions and prediction intervals for the cumulative number of failures, over a range of time, for the overall fleet of transformers.

## 1. Introduction.

1.1. *Background.* Electrical transmission is an important part of the US energy industry. There are approximately 150,000 high-voltage power transmission transformers in service in the US. Unexpected failures of transformers can cause large economic losses. Thus, prediction of remaining life of transformers is an important issue for the owners of these assets. The prediction of the remaining life can be based on historical lifetime information about the transformer population (or fleet). However, because the lifetimes of some transformers extend over several decades, transformer lifetime data are complicated.

This paper describes the analysis of transformer lifetime data from an energy company. Based on the currently available data, the company wants to know the remaining life of the healthy individual transformers in its fleet and the rate at which these transformers will fail over time. To protect sensitive and proprietary information, we will not use the name of the company. We also code the name of the transformer manufacturers and modify the serial numbers of the transformers in the data. We use a parametric lifetime model to describe the lifetime distribution of individual transformers. We present a statistical procedure for computing a prediction interval for remaining life for individuals and for the cumulative number failing in the future.

The energy company began careful archival record keeping in 1980. The dataset provided to us contains complete information on all units that were installed after 1980 (i.e., the installation dates of all units and date of failure for those that failed). We also have information on units that were installed before January 1, 1980 and failed *after* January 1, 1980. We do not, however, have any information on units installed and failed before 1980. Thus, transformers that were installed before 1980 must be viewed as transformers sampled from truncated distribution(s). Units that are still in service have lifetimes that are right censored. Hence, the data are left truncated and right censored. For those units that are left truncated or right censored (or both), the truncation times and censoring times differ from unit-to-unit because of the staggered entry of the units into service.

There are standard statistical methods for estimating distribution parameters with truncated data described, for example, in Meeker and Escobar (2003) and Meeker and Escobar [(1998), Chapter 11], but such methods appear not to be available in commercial software. Meeker and Escobar (2008), a free package for reliability data analysis, does allow for truncated data. Most of the computations needed to complete this paper, however, required extending this software.

In this paper we outline a general methodology for reliability prediction in complicated situations that involve the need for dealing with stratification, truncation, and censoring. In addition to describing our approach for dealing with these complications, we show how to produce calibrated prediction intervals by using the random weighted bootstrap and an approximation based on a refined central limit theorem.

1.2. *A general approach to statistical prediction of transformer life.* Our approach to the prediction problem will be divided into the following steps.

1. *Stratification:* A simple lifetime model fit to a pooled mixture of disparate populations can lead to incorrect conclusions. For example, engineering knowledge suggests that there is an important difference between old transformers and new transformers because old transformers were over-engineered. Thus, we first stratify all transformers into relatively homogeneous groups that have similar lifetime distributions. This grouping will be based on manufacturer and date of installation. The groupings will be determined from a combination of knowledge of transformer failure mechanisms, manufacturing history, and data analysis. Each group will have its own set of parameters. The parameters will be estimated from the available lifetime data by using the maximum likelihood (ML) method. We may, however, be able to reduce the number of parameters needed to be estimated by, for example, assuming a common shape parameter across some of the groups (from physics of failure, we know that similar failure modes can often be expected to be described by distributions with similar shape parameters).

2. *Lifetime distribution:* Estimate the lifetime probability distribution for each group of transformers from the available lifetime data.

3. *Remaining life distribution:* Identify all transformers that are at risk to fail (the "risk set"). Each of these transformers belongs to one of the above-mentioned groups of transformers. For each transformer in the risk set, compute an estimate of the distribution of remaining life (this is the conditional distribution of remaining life, given the age of the individual transformer).

4. *Expected number of transformers failing:* Having the distribution of remaining life on each transformer that is at risk allows the computation of the estimated expected number of transformers failing in each future interval of time (e.g., future months). We use this estimated expected number failing as a prediction of population behavior.

5. *Prediction intervals:* It is also important to compute prediction intervals to account for the statistical uncertainty in the predictions (statistical uncertainty accounts for the uncertainty due to the limited sample size and the variability in future failures, but assumes that the statistical model describing transformer life is correct).

6. *Sensitivity analysis:* To compute our predictions, we need to make assumptions about the stratification and lifetime distributions. There is not enough information in the data or from the engineers at the company to be certain that these assumptions are correct. Thus, it is important to perturb the assumptions to assess their effect on answers.

1.3. *Overview.* The rest of the paper is organized as follows. Section 2 describes our exploratory analysis of the transformer lifetime data and several potentially important explanatory variables. Section 3 describes the model and methods for estimating the transformer lifetime distributions. Section 4 gives details on stratifying the data into relatively homogeneous groups and our regression analyses. Section 5 shows how estimates of the transformer lifetime distributions lead to age-adjusted distributions of remaining life for individual transformers and how these distributions can be used as a basis for computing a prediction interval for remaining life for individual transformers. Section 6 provides predictions for the cumulative number of failures for the overall population of transformers now in service, as a function of time. Section 7 presents sensitivity analysis on the prediction results. Section 8 concludes with some discussion and describes areas for future research.

TABLE 1
*Summary of the number of failed, censored, and truncated units for the different manufacturers*

| **Manufacturer** | **Failed** | **Censored** | **Truncated** | **Total** |
|---|---|---|---|---|
| MA | 9 | 37 | 0 | 46 |
| MB | 6 | 44 | 49 | 50 |
| MC | 23 | 127 | 122 | 150 |
| MD | 6 | 22 | 27 | 28 |
| ME | 9 | 150 | 137 | 159 |
| Other | 9 | 268 | 106 | 277 |

## 2. The transformer lifetime data.

The dataset used in our study contains 710 observations with 62 failures. Table 1 gives a summary of the number of failed, censored, and truncated units for the different manufacturers. Figure 1 is an event plot of a systematic subset of the data.

### 2.1. *Failure mechanism.*

Transformers, for the most part, fail when voltage stress exceeds the dielectric strength of the insulation. The insulation in a transformer is made of a special kind of paper. Over time, the paper will chemically degrade, leading to a loss in dielectric strength, and eventual failure. The rate of degradation depends primarily on operating temperature. Thus, all other things being equal, transformers that tend to run at higher load, with correspondingly higher temperatures, would be expected to fail sooner than those running at lower loads. Events such as short circuits on the transmission grid can cause momentary thermal spikes that can be especially damaging to the insulation.

### 2.2. *Early failures.*

Seven units failed within the first 5 years of installation. The lifetimes for these units are short compared with the vast majority of units that failed or will fail with age greater than 10 years. These early failures are believed to have been due to a defect related failure mode that is different from all of the other failures. The inclusion of these early failures in the analysis leads to an indication of an approximately constant hazard function for transformer life, which is inconsistent with the known predominant aging failure mode. Thus, we considered these early failures to be right censored at the time of failure. This is justified because the primary goal of our analysis is to model the failure mode for the future failures for the remaining units. It is reasonable to assume that there are no more defective units in the population for which predictions are to be generated.

2.3. *Explanatory variables.* Engineering knowledge suggests that the insulation type and cooling classes may have an effect on the lifetime of transformers. Thus, the effects that these two variables have on lifetime are studied in this paper.

*Insulation.* The transformers are rated at either a 55 or 65 *degree rise.* This variable defines the average temperature rise of the winding, above ambient, at which the transformer can operate in continuous service. For example, a 55 degree-rise rated transformer operated at a winding temperature of 95 degrees should, if the engineering model describing this phenomena is adequate, have the same life as a 65 degree-rise rated transformer operated at a winding temperature of 105 degrees. The two categories of the insulation class are denoted by “d55” and “d65,” respectively.

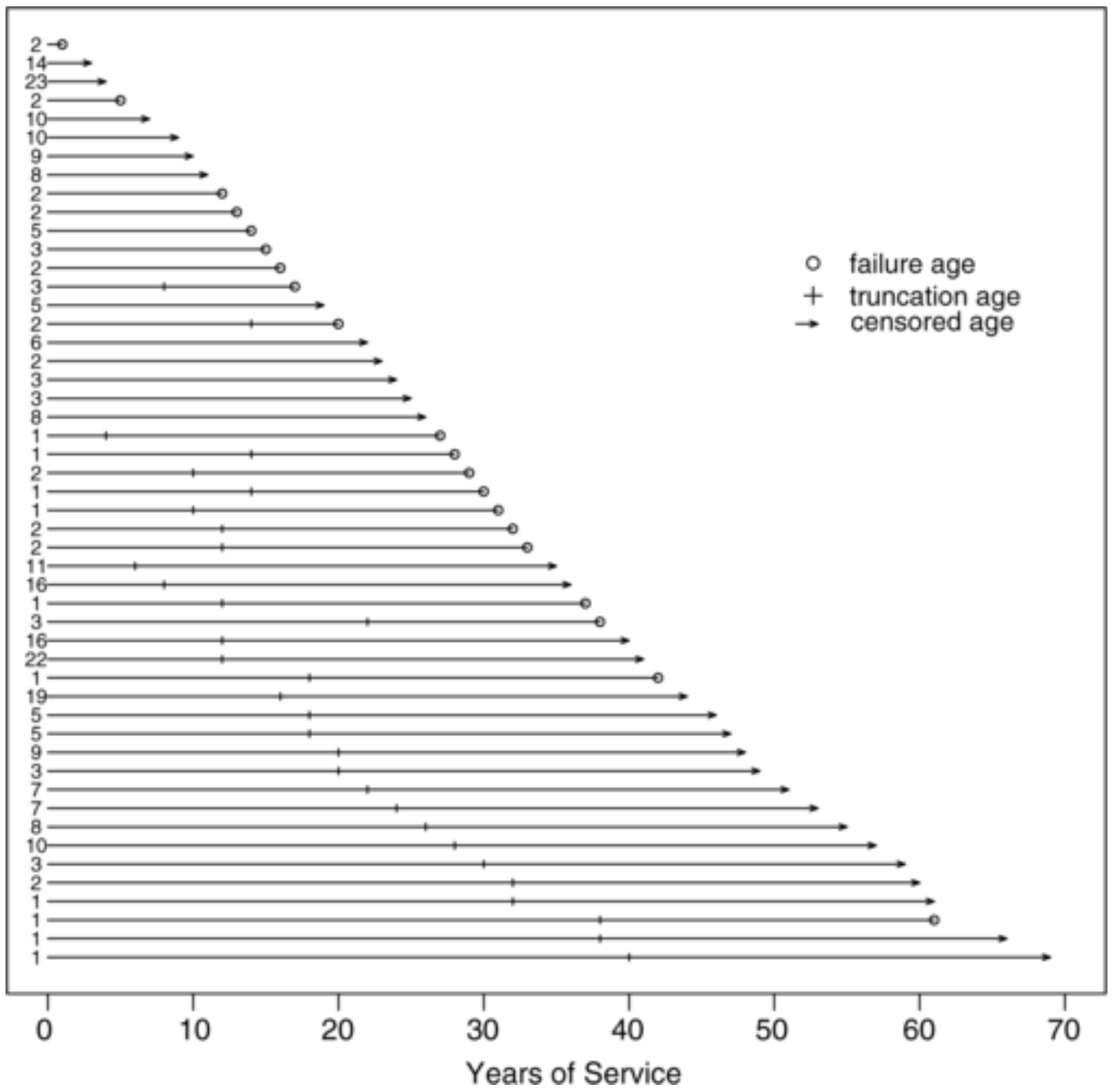

FIG. 1. *Service-time event plot of a systematic subset of the transformer lifetime data. The numbers in the left panel of the plot are counts for each line.*

*Cooling.* A transformer's cooling system consists of internal and external subsystems. The internal subsystem uses either natural or forced flow of oil. Forced flow is more efficient. The external cooling system uses either air or water cooling. Water cooling is more efficient. The external cooling media circulation is again either natural or forced. Forced circulation is usually used on larger units and is more efficient but is activated only when the temperature is above a certain threshold. The cooling methods for the transformers in the data are categorized into four groups: natural internal oil and natural external air/water (NINE), natural internal oil and forced external air/water (NIFE), forced internal oil and forced external air/water (FIFE), and unknown.

## 3. Statistical lifetime model for left truncated and right censored data.

3.1. *The lifetime model.* We denote the lifetime of a transformer by $T$ and model this time with a log-location-scale distribution. The most commonly used distributions for lifetime, the Weibull and lognormal, are members of this family. The cumulative distribution function (c.d.f.) of a log-location-scale distribution can be expressed as

$$F(t;\mu,\sigma)=\Phi\left[\frac{\log(t)-\mu}{\sigma}\right],$$

where $\Phi$ is the standard c.d.f. for the location-scale family of distributions (location 0 and scale 1), $\mu$ is the location parameter, and $\sigma$ is the scale parameter. The corresponding probability density function (p.d.f.) is the first derivative of the c.d.f. with respect to time and is given by

$$f(t;\mu,\sigma)=\frac{1}{\sigma t}\phi\left[\frac{\log(t)-\mu}{\sigma}\right],$$

where $\phi$ is the standard p.d.f. for the location-scale family of distributions. The hazard function is $h(t;\mu,\sigma)=f(t;\mu,\sigma)/[1-F(t;\mu,\sigma)]$. For the lognormal distribution, replace $\Phi$ and $\phi$ above with $\Phi_{\mathrm{nor}}$ and $\phi_{\mathrm{nor}}$, the standard normal c.d.f. and p.d.f., respectively. The c.d.f. and p.d.f. of the Weibull random variable $T$ are

$$F(t;\mu,\sigma)=\Phi_{\mathrm{sev}}\left[\frac{\log(t)-\mu}{\sigma}\right]\quad\text{and}\quad f(t;\mu,\sigma)=\frac{1}{\sigma t}\phi_{\mathrm{sev}}\left[\frac{\log(t)-\mu}{\sigma}\right],$$

where $\Phi_{\mathrm{sev}}(z)=1-\exp[-\exp(z)]$ and $\phi_{\mathrm{sev}}(z)=\exp[z-\exp(z)]$ are the standard (i.e., $\mu=0,\sigma=1$) smallest extreme value c.d.f. and p.d.f., respectively.

The c.d.f. and p.d.f. of the Weibull random variable $T$ can also be expressed as

$$F(t;\eta,\beta) = 1 - \exp\left[-\left(\frac{t}{\eta}\right)^{\beta}\right] \quad \text{and} \quad f(t;\eta,\beta) = \left(\frac{\beta}{\eta}\right)\left(\frac{t}{\eta}\right)^{\beta-1} \exp\left[-\left(\frac{t}{\eta}\right)^{\beta}\right],$$

where $\eta = \exp(\mu)$ is the scale parameter and $\beta = 1/\sigma$ is the shape parameter. If the Weibull shape parameter $\beta > 1$, the Weibull hazard function is increasing (corresponding to wearout); if $\beta = 1$, the hazard function is a constant; and if $\beta < 1$, the hazard function is decreasing. The location-scale parametrization is, however, more convenient for regression analysis.

3.2. *Censoring and truncation.* Right-censored lifetime data result when unfailed units are still in service (unfailed) when data are analyzed. A transformer still in service in March 2008 (the "data-freeze" point) is considered as a censored unit in this study.

Truncation, which is similar to but different from censoring, arises when failure times are observed only when they take on values in a particular range. When the existence of the unseen "observation" is not known for observations that fall outside the particular range, the data that are observed are said to be truncated. Because we have no information about transformers that were installed *and failed* before 1980, the units that were installed before 1980 and failed after 1980 should be modeled as having been sampled from a left-truncated distribution(s). Ignoring truncation causes bias in estimation.

3.3. *Maximum likelihood estimation.* Let $t_i$ denote the lifetime or survival time of transformer $i$, giving the number of years of service between the time the transformer was installed until it failed (for a failed transformer) or until the data-freeze point (for a surviving transformer). Here, $i = 1, \ldots, n$, where $n$ is the number of transformers in the dataset. Let $\tau_i^L$ be the left truncation time, giving the time at which the life distribution of transformer $i$ was truncated on the left. More precisely, $\tau_i^L$ is the number of years between the transformer's manufacturing date and 1980 for transformers installed before 1980. Let $\nu_i$ be the truncation indicator. In particular, $\nu_i = 0$ if transformer $i$ is truncated (installed before 1980) and $\nu_i = 1$ if transformer $i$ is not truncated (installed after 1980). Let $c_i$ be the censoring time (time that a transformer has survived) and let $\delta_i$ be the censoring indicator. In particular, $\delta_i = 1$ if transformer $i$ failed and $\delta_i = 0$ if it was censored (not yet failed).

The likelihood function for the transformer lifetime data is

$$\begin{aligned} L(\boldsymbol{\theta}|DATA) = \prod_{i=1}^{n} f(t_i;\boldsymbol{\theta})^{\delta_i \nu_i} \times \left[\frac{f(t_i;\boldsymbol{\theta})}{1 - F(\tau_i^L;\boldsymbol{\theta})}\right]^{\delta_i(1-\nu_i)} \\ \times [1 - F(c_i;\boldsymbol{\theta})]^{(1-\delta_i)\nu_i} \times \left[\frac{1 - F(c_i;\boldsymbol{\theta})}{1 - F(\tau_i^L;\boldsymbol{\theta})}\right]^{(1-\delta_i)(1-\nu_i)}. \end{aligned} \tag{1}$$

Here $\boldsymbol{\theta}$ is a vector that gives the location parameter ($\mu_i$) and scale parameters ($\sigma_i$) for each transformer. The exact structure of $\boldsymbol{\theta}$ depends on the context of the model. For example, in Section 4.1, we stratify the data into $J$ groups with $n_j$ transformers in group $j$ and fit a single distribution to each group. For this model we assume that observations from group $j$ have the same location ($\mu_j$) and scale parameters ($\sigma_j$). Thus,

$$\boldsymbol{\theta} = (\underbrace{\mu_1, \ldots, \mu_1}_{\text{Group } 1}, \ldots, \underbrace{\mu_J, \ldots, \mu_J}_{\text{Group } J}, \underbrace{\sigma_1, \ldots, \sigma_1}_{\text{Group } 1}, \ldots, \underbrace{\sigma_J, \ldots, \sigma_J}_{\text{Group } J})'.$$

For notational simplicity, we also use $F(t_i; \boldsymbol{\theta}) = F(t_i; \mu_i, \sigma_i)$ and $f(t_i; \boldsymbol{\theta}) = f(t_i; \mu_i, \sigma_i)$. In our regression models, $\mu_i$ may depend on the values of the explanatory variables. The ML estimate $\widehat{\boldsymbol{\theta}}$ is obtained by finding the values of the parameters that maximize the likelihood function in (1).

## 4. Stratification and regression analysis.

4.1. *Stratification.* As described in Section 1.2, we need to stratify the data into relatively homogeneous groups. Manufacturer and installation year were used as preliminary stratification variables. The choice of installation year as the stratification variable is strongly motivated by the design change of transformers. There is a big difference between the old transformers and new transformers. The engineers indicate that old transformers were over-engineered and can last a long time. For example, there are transformers installed in 1930s that are still in service, as shown in Figure 1. Due to the competition in the transformer manufacture industry and the need of reducing manufacturing costs, the new transformers are not as “strong” as old ones.

The transformers manufactured by the same manufacturer were divided into two groups (New and Old) based on age (installation year). We chose the cutting year for this partitioning to be 1987. In Section 7.1 we give the results of a sensitivity analysis that investigated the effects of changing the cutting year. There are only one or two failures in some groups (i.e., MC_New, ME_New, and Other_New). These groups were combined together as MC.ME.Other_New. Note that all MA units were installed after 1990 and all MB units were installed before 1987.

Figure 2 is a multiple Weibull probability plot showing the nonparametric and the Weibull ML estimates of the c.d.f. for all of the individual groups. The nonparametric estimates (those points in Figure 2) are based on the method for truncated/censored data described in Turnbull (1976). The points in Figure 2 were plotted at each observed lifetime (censored units were not plotted) and at the midpoint of the step of the Turnbull c.d.f. estimates, as suggested in Meeker and Escobar [(1998), Section 6.4.2] and

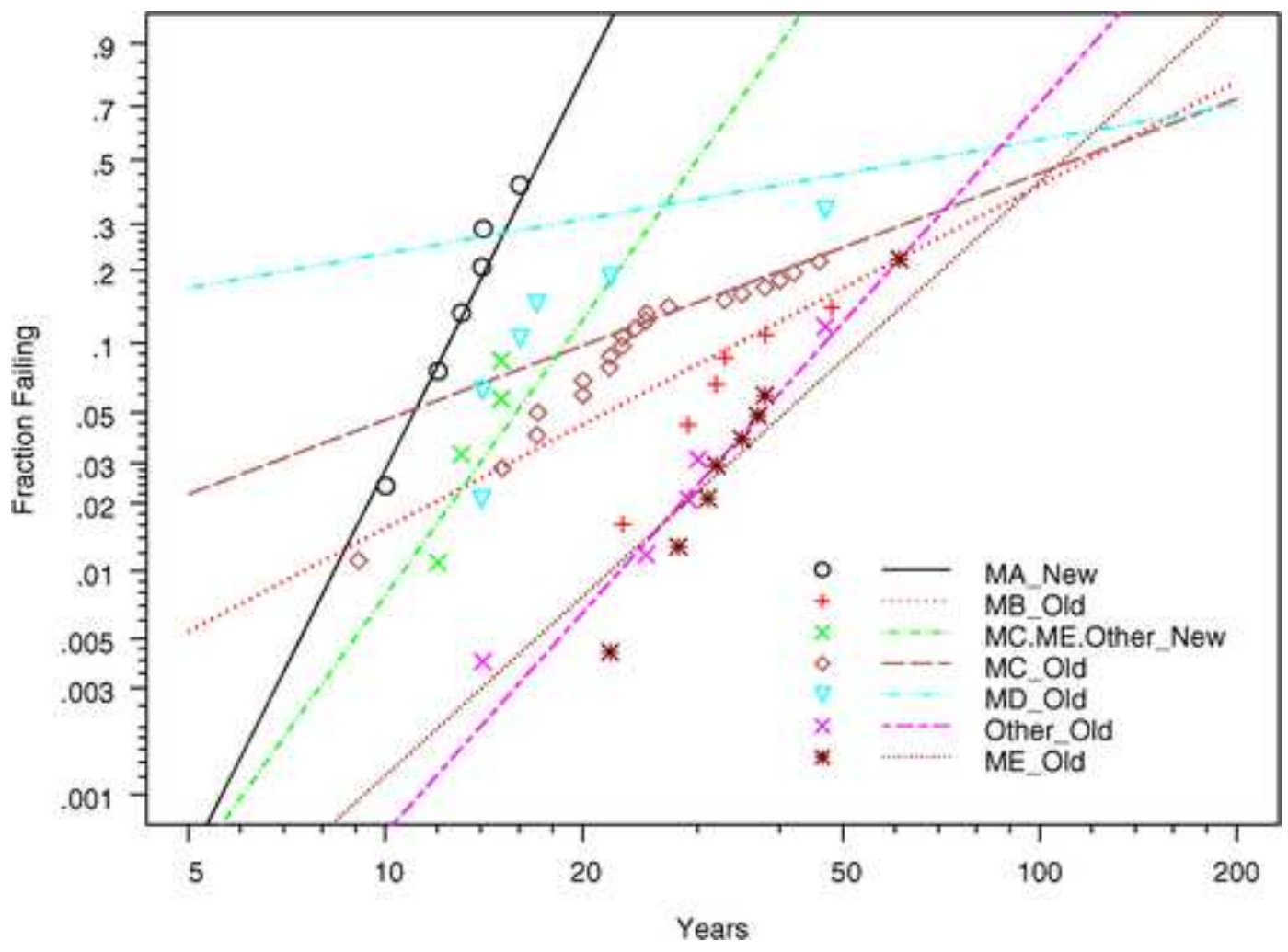

Fig. 2. *Weibull probability plot with the ML estimates of the c.d.f.s for each of the individual groups*

Lawless [(2003), Section 3.3]. Table 2 gives the ML estimates and standard errors of the Weibull distribution parameters for each group.

Note that the nonparametric and the parametric estimates in Figure 2 do not agree well for the Old groups. This is due to the truncation in these groups. When sampling from a truncated distribution, the ML estimator based on the likelihood in (1) is consistent. The nonparametric estimator used in the probability plots, however, is not consistent if all observations are truncated. Because almost all of the observations are truncated in the Old groups, we would not expect the parametric and nonparametric estimates to agree well, even in moderately large finite samples.

Based on the ML estimates for the individual groups, the dataset was partitioned into two large groups: the Old group with slowly increasing hazard rate ($\widehat{\beta} \approx 2$), and the New group with a more rapidly increasing hazard rate ($\widehat{\beta} \approx 5$). The Old group consists of MB_Old, MC_Old, Other_Old, and ME_Old, and the New group consists of MA_New and MC.ME.Other_New. When we do regression analyses in Section 4.4, we assume that there is a common shape parameter for all of the transformers in the *Old group* and a different common shape parameter for all of the transformers in the *New group*. This assumption is supported by the lifetime data, as can be seen in Figure 2, and by doing likelihood ratio tests (details not given here).

4.2. *Distribution choice.* We also fit individual lognormal distributions and made a lognormal probability plot (not shown here) that is similar to Figure 2. Generally, the Weibull distributions fit somewhat better, both visually in the probability plot and in terms of the loglikelihood values of the ML estimates. There is a physical/probabilistic explanation for this conclusion. In the transformer, there are many potential locations where the voltage stress could exceed the dielectric stress. The transformer will fail the first time such an event occurs. That is, a transformer's lifetime is controlled by the distribution of a minimum. The Weibull distribution is one of the limiting distribution of minima.

4.3. *A problem with the MD group data.* As shown in Table 2, the estimate of the Weibull shape parameter for the MD group is $\widehat{\beta} = 0.51$, implying a strongly decreasing hazard function. Such a decreasing hazard is not consistent with the known aging failure mode of the transformer insulation. This problem with the estimation is caused by the extremely heavy truncation. More details about this estimation problem are available in the supplemental article [Hong, Meeker, and McCalley (2009)]. As a remedy, in the estimation and modeling stage, we exclude the MD units. When we make the predictions, however, we include the MD_Old units that are currently in service in the Old group and the single MD_New unit in the New group based on engineering knowledge about the designs.

4.4. *Regression analysis.* In this section we extend the single distribution models fit in Section 4.1 to regression models. For details on parametric regression analysis for lifetime data, see, for example, Lawless (2003) or Meeker and Escobar [(1998), Chapter 17]. In our models, the location parameter $\mu$ is treated as a function of explanatory variable $\mathbf{x}$, denoted by $\mu(\mathbf{x}) = g(\mathbf{x}, \boldsymbol{\beta})$, where $\mathbf{x} = (x_1, x_2, \ldots, x_p)'$ and $\boldsymbol{\beta} = (\beta_0, \beta_1, \ldots, \beta_p)'$. In the case of linear regression $g(\mathbf{x}, \boldsymbol{\beta}) = \mathbf{x}'\boldsymbol{\beta}$.

In the next two sections we fit separate regression models for the strata identified in Section 4.1. The explanatory variables considered in the regression modeling are `Manufacturer`, `Insulation`, and `Cooling`, all of which are categorical variables.

TABLE 2
*Weibull ML estimates of parameters and standard errors for each group*

| **Group** | $\widehat{\eta}$ | $\widehat{se}_{\widehat{\eta}}$ | $\widehat{\beta}$ | $\widehat{se}_{\widehat{\beta}}$ | **Failures** | **Total** |
|---|---|---|---|---|---|---|
| MA_New | 18.39 | 1.607 | 5.83 | 1.796 | 6 | 46 |
| MC.ME.Other_New | 32.75 | 8.920 | 4.09 | 1.594 | 4 | 167 |
| MB_Old | 150.27 | 97.953 | 1.54 | 1.057 | 6 | 50 |
| MC_Old | 157.81 | 61.187 | 1.10 | 0.381 | 20 | 133 |
| MD | 136.81 | 109.638 | 0.51 | 0.499 | 6 | 28 |
| Other_Old | 93.49 | 36.751 | 3.26 | 1.288 | 5 | 137 |
| ME_Old | 124.85 | 44.351 | 2.66 | 0.952 | 8 | 149 |

*The Old group.* Table 3 compares the loglikelihood values for the Weibull regression models fit to the Old group. Likelihood ratio tests show that `Manufacturer` and `Insulation` are not statistically important (i.e., the values of the loglikelihood for Models 2 and 3 are only slightly larger then that for Model 1). Hence, the final model for the Old group is $\mu(\mathbf{x}) = \texttt{Cooling}$. Table 4 gives ML estimates and confidence intervals for parameters for the final model for the Old group. Figure 3a gives the Weibull probability plot showing the Weibull regression estimate of the c.d.f.s for the different cooling categories. The slopes of the fitted lines are the same because of the constant shape parameter assumption in our model.

*The New group.* Table 5 compares the loglikelihood values for the Weibull regression models fit to the New group. `Insulation` is not in the model because it only has one level in the New group. Likelihood ratio tests show that `Manufacturer` is statistically important. Hence, the final model for the New group is $\mu(\mathbf{x}) = \texttt{Manufacturer}$. Table 6 gives ML estimates and confidence intervals for the final regression model parameters for the New group. Figure 3b is a Weibull probability plot showing the ML estimates of the c.d.f.s for the two manufacturers in this group.

**5. Predictions for the remaining life of individual transformers.** In this section we develop a prediction interval procedure to capture, with $100(1-\alpha)\%$ confidence, the future failure time of an individual transformer, conditional on survival until its present age, $t_i$. The prediction interval is denoted

TABLE 3
*Model comparison for the Old group based on the Weibull distribution*

| Model | | Loglikelihood |
|---|---|---|
| 1 | $\mu(\mathbf{x}) = \texttt{Cooling}$ | −103.663 |
| 2 | $\mu(\mathbf{x}) = \texttt{Manufacturer} + \texttt{Cooling}$ | −100.268 |
| 3 | $\mu(\mathbf{x}) = \texttt{Manufacturer} + \texttt{Cooling} + \texttt{Insulation}$ | −100.198 |

TABLE 4
*Weibull ML estimates and confidence intervals for the Old group*

| Parameter | MLE | Std. err. | 95% lower | 95% upper |
|---|---|---|---|---|
| $\widehat{\eta}$(NIFE) | 127.22 | 25.112 | 86.401 | 187.317 |
| $\widehat{\eta}$(FIFE) | 92.66 | 17.305 | 64.251 | 133.607 |
| $\widehat{\eta}$(NINE) | 346.47 | 186.249 | 120.808 | 993.665 |
| $\widehat{\eta}$(Unknown) | 32.12 | 4.750 | 24.042 | 42.927 |
| $\widehat{\beta}$ | 2.22 | 0.357 | 1.624 | 3.045 |

Table 5

*Model comparison for the New group based on the Weibull distribution*

| **Model** | | **Loglikelihood** |
|---|---|---|
| 4 | $\mu(\mathbf{x}) = \mu$ | −25.268 |
| 5 | $\mu(\mathbf{x}) = \texttt{Manufacturer}$ | −20.138 |
| 6 | $\mu(\mathbf{x}) = \texttt{Manufacturer} + \texttt{Cooling}$ | −18.089 |

by $[\underset{\sim}{T}_i, \widetilde{T}_i]$. The c.d.f. for the lifetime of a transformer, conditional on surviving until time $t_i$, is

$$F(t|t_i;\boldsymbol{\theta}) = \Pr(T \le t|T > t_i) = \frac{F(t;\boldsymbol{\theta}) - F(t_i;\boldsymbol{\theta})}{1 - F(t_i;\boldsymbol{\theta})}, \qquad t \ge t_i. \tag{2}$$

This conditional c.d.f. provides the basis of our predictions and prediction intervals.

5.1. *The naive prediction interval procedure.* A simple naive prediction interval procedure (also known as the "plug-in" method) provides an approximate interval that we use as a start toward obtaining a more refined interval. The procedure simply takes the ML estimates of the parameters and substitutes them into the estimated conditional probability distributions in (2) (one distribution for each transformer). The estimated probability distributions can then be used as a basis for computing predictions and prediction intervals. Let $100(1-\alpha)\%$ be the nominal coverage probability. The *coverage probability* is defined as the probability that the prediction interval procedure will produce an interval that captures what it is intended to capture.

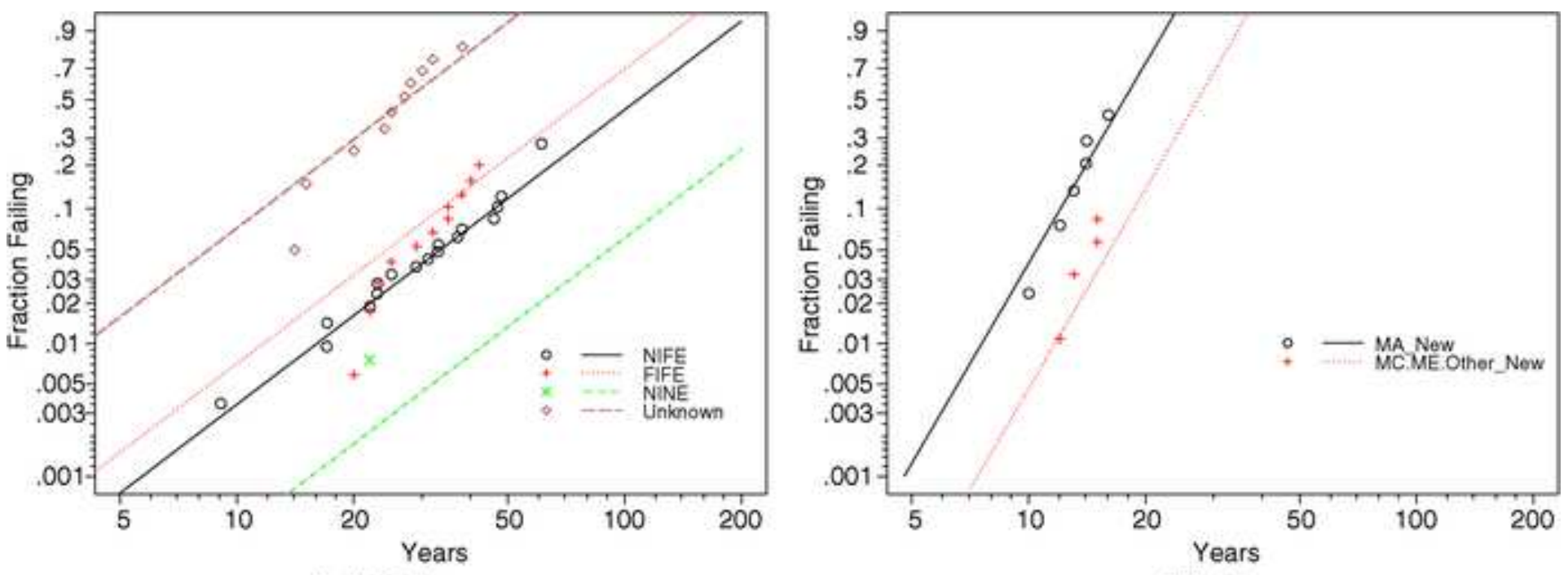

(a) Old group (b) New group

Fig. 3. *Weibull probability plots showing the ML estimates of the c.d.f.s for the Old group and the New group regression models.*

The naive $100(1-\alpha)\%$ prediction interval for a transformer having age $t_i$ is $[\underset{\sim}{T}_i, \widetilde{T}_i]$, where $\underset{\sim}{T}_i$ and $\widetilde{T}_i$ satisfy $F(\underset{\sim}{T}_i|t_i, \widehat{\boldsymbol{\theta}}) = \alpha_{\mathrm{l}}, F(\widetilde{T}_i|t_i, \widehat{\boldsymbol{\theta}}) = 1 - \alpha_{\mathrm{u}}$. Here $\alpha_{\mathrm{l}}$ and $\alpha_{\mathrm{u}}$ are the lower and upper tail probabilities, respectively, and $\alpha_{\mathrm{l}} + \alpha_{\mathrm{u}} = \alpha$. We choose $\alpha_{\mathrm{l}} = \alpha_{\mathrm{u}} = \alpha/2$. This simple procedure ignores the uncertainty in $\widehat{\boldsymbol{\theta}}$. Thus, the interval coverage probability of this simple procedure is generally smaller than the nominal confidence level. The procedure needs to be calibrated so that it will have a coverage probability that is closer to the nominal confidence level.

5.2. *Calibration of the naive prediction interval.* Calibration of the naive prediction interval procedure to account for statistical uncertainty can be done through asymptotic expansions [Komaki (1996), Barndorff-Nielsen and Cox (1996)] or by using Monte Carlo simulation/bootstrap re-sampling methods [Beran (1990), Escobar and Meeker (1999)]. Lawless and Fredette (2005) show how to use a predictive distribution approach that provides intervals that are the same as the calibrated naive prediction interval.

In practice, simulation is much easier and is more commonly used to calibrate naive prediction interval procedures. In either case, the basic idea is to find an input value for the coverage probability (usually larger than the nominal value) that gives a procedure that has the desired nominal coverage probability. In general, the actual coverage probability of a procedure employing calibration is still only approximately equal to the nominal confidence level. The calibrated procedure, if it is not exact (i.e., actual coverage probability is equal to the nominal), can be expected to provide a much better approximation than the naive procedure.

5.3. *The random weighted bootstrap.* Discussion of traditional bootstrap resampling methods for lifetime/survival data can be found, for example, in Davison and Hinkley (1997). Due to the complicated data structure and sparsity of failures over the combinations of different levels of explanatory variables, however, the traditional bootstrap method is not easy to implement and may not perform well. Bootstrapping with the commonly used simple random sampling with replacement with heavy censoring can be problematic, as it can result in bootstrap samples without enough failures for the estimation of parameters (only about 9% of the transformers had failed). A parametric bootstrap would require distribution assumptions on the truncation time and censoring time and this information is not available. The stratification, regression modeling, and especially the left truncation lead to other difficulties with bootstrapping. The random weighted likelihood bootstrap procedure, introduced by Newton and Raftery (1994), provides a versatile, effective, and easy-to-use method to generate bootstrap samples for such more complicated problems. The procedure uses the following steps:

1. Simulate random values $Z_i, i = 1, 2, \ldots, n$, that are i.i.d. from a distribution having the property $\mathrm{E}(Z_i) = [\mathrm{Var}(Z_i)]^{1/2}$.
2. The random weighted likelihood is $L^*(\boldsymbol{\theta}|DATA) = \prod_{i=1}^{n}[L_i(\boldsymbol{\theta}|DATA)]^{Z_i}$, where $L_i(\boldsymbol{\theta}|DATA)$ is the likelihood contribution from an individual observation.
3. Obtain the ML estimate $\widehat{\boldsymbol{\theta}}^*$ by maximizing $L^*(\boldsymbol{\theta}|DATA)$.
4. Repeat steps 1–3 $B$ times to get $B$ bootstrap samples $\widehat{\boldsymbol{\theta}}_b^*, b = 1, 2, \ldots, B$.

TABLE 6
*Weibull ML estimates and confidence intervals for the New group*

| **Parameter** | **MLE** | **Std. err.** | **95% lower** | **95% upper** |
|---|---|---|---|---|
| $\widehat{\eta}$(MA_New) | 18.94 | 1.850 | 15.641 | 22.936 |
| $\widehat{\eta}$(MC.ME.Other_New) | 29.29 | 4.548 | 21.602 | 39.706 |
| $\widehat{\beta}$ | 5.01 | 1.229 | 3.098 | 8.104 |

Barbe and Bertail [(1995), Chapter 2] discuss how to choose the random weights by using an Edgeworth expansion. Jin, Ying, and Wei (2001) showed that the distribution of $\sqrt{n}(\widehat{\boldsymbol{\theta}}^* - \widehat{\boldsymbol{\theta}})$ (given the original data) can be used to approximate the distribution of $\sqrt{n}(\widehat{\boldsymbol{\theta}} - \boldsymbol{\theta})$, if one uses i.i.d. positive random weights generated from continuous distribution with $\mathrm{E}(Z_i) = [\mathrm{Var}(Z_i)]^{1/2}$. They pointed out that the resampling method is rather robust for different choices of the distribution of $Z_i$, under this condition. We used $Z_i \sim \mathrm{Gamma}(1, 1)$ in this paper. We also tried alternative distributions, such as the Gamma$(1, 0.5)$, Gamma$(1, 2)$, and Beta$(\sqrt{2} - 1, 1)$. The resulting intervals were insensitive to the distribution used, showing similar robustness for our particular application.

5.4. *Calibrated prediction intervals.* For an individual transformer with age $t_i$, the calibrated prediction interval of remaining life can be obtained by using the following procedure. Lawless and Fredette's predictive distribution [Lawless and Fredette (2005)] are used here:

1. Simulate $T_{ib}^*, b = 1, \ldots, B$, from distribution $F(t|t_i, \widehat{\boldsymbol{\theta}})$.
2. Compute $U_{ib}^* = F(T_{ib}^*|t_i, \widehat{\boldsymbol{\theta}}_b^*), b = 1, \ldots, B$.
3. Let $u_i^{\mathrm{l}}, u_i^{\mathrm{u}}$ be, respectively, the lower and upper $\alpha/2$ sample quantiles of $U_{ib}^*, b = 1, \ldots, B$. The $100(1 - \alpha)\%$ calibrated prediction interval can be obtained by solving for $\underset{\sim}{T}_i$ and $\widetilde{T}_i$ in $F(\underset{\sim}{T}_i|t_i\widehat{\boldsymbol{\theta}}) = u_i^{\mathrm{l}}$ and $F(\widetilde{T}_i|t_i, \widehat{\boldsymbol{\theta}}) = u_i^{\mathrm{u}}$, respectively.

5.5. *Prediction results.* In this section we present prediction intervals for the remaining life for individual transformers based on using the Weibull distribution and a stratification cutting at year 1987. Figure 4 shows 90% prediction intervals for remaining life for a subset of individual transformers that are at risk. The Years axis is logarithmic.

There are some interesting patterns in these results. In particular, for a group of relatively young transformers in the same group (young relative to expected life) and with the same values of the explanatory variable(s), the prediction intervals are similar (but not exactly the same because of the conditioning on actual age). For a unit in such a group (i.e., one that has been in service long enough to have its age fall within the prediction intervals for the younger units), however, the lower endpoints of the interval are very close to the current age of the unit. Intervals for such units can be rather short, indicating that, according to our model, they are at high risk to failure. See, for example, unit MA_New200 in Figure 4. Interestingly, as we were finishing this work, we learned of a recent failure of a transformer that had such a prediction interval.

Units, like MA_New200, that are predicted to be at especially high risk for failure in the near term are sometimes outfitted with special equipment to continuously (hourly) monitor, communicate, and archive transformer condition measurements that are useful for detecting faults that may lead to failure. These measurements are taken from the transformer insulating oil and most commonly indicate the presence of dissolved gases but also may indicate other attributes, including moisture content and loss of dielectric strength. Dissolved gas analysis (DGA) is automatically-performed by these monitors and is important in the transformer maintenance process, because it can be used to predict anomalous and dangerous conditions such as winding overheating, partial discharge, or arcing in the transformer. Without such a monitor, DGA is performed by sending an oil sample to a laboratory. These lab tests are routinely performed on a 6–12 month basis for healthy transformers but more frequently if a test indicates a potential problem. If an imminent failure can be detected early enough, the transformer can be operated under reduced loading until replaced, to avoid costly catastrophic failures that sometimes cause explosions. Lab testing, although generally useful, exposes the transformer to possible rapidly deteriorating failure conditions between tests. Continuous monitoring eliminates this exposure but incurs the investment price of the monitoring equipment. Although this price is typically less than 1% of the transformer cost, the large number of transformers in a company's fleet prohibits monitoring of all of them.

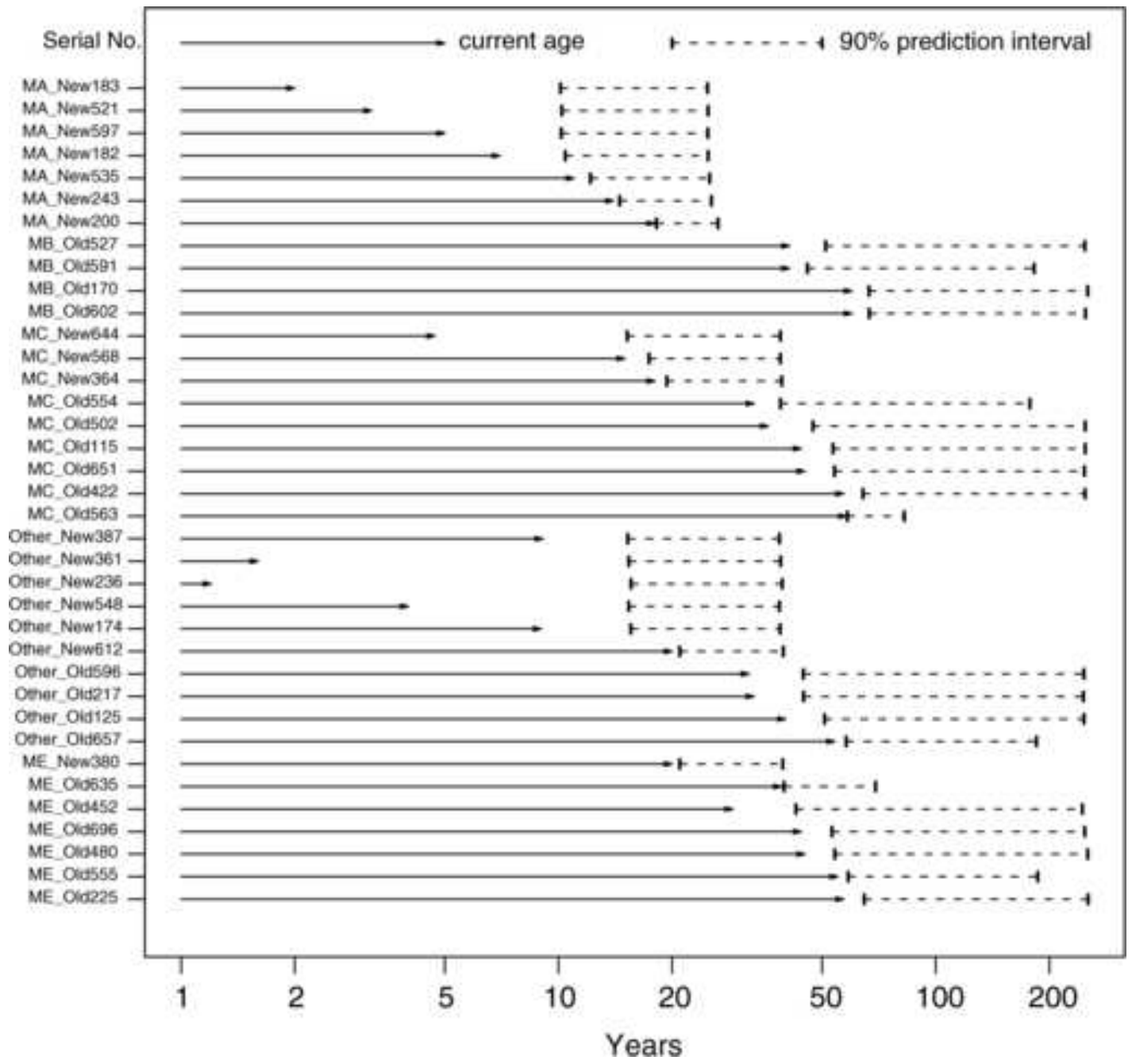

Fig. 4. *Weibull distribution 90% prediction intervals for remaining life for a subset of individual at-risk transformers.*

**6. Prediction for the cumulative number of failures for the population.** This section describes a method for predicting the cumulative number of future failures in the population, as a function of time. For the population of transformers, we will predict the cumulative number of failures by the end of each month for the next 10 years. We also compute corresponding calibrated pointwise prediction intervals, quantifying the statistical uncertainty and failure process variability. Such predictions and intervals are needed for planning of capital expenditures.

6.1. *Population prediction model.* From (2), for an individual transformer that has survived and has age $t_i$ at the data-freeze time, the conditional probability of failure between age $t_i$ and a future age $t_i^w$ (the amount of time in service for transformer $i$ at a specified date in the future) is $\rho_i = F(t_i^w|t_i, \boldsymbol{\theta})$. The ML estimator of $\rho_i$ is $\widehat{\rho}_i = F(t_i^w|t_i, \widehat{\boldsymbol{\theta}})$. Note that the times $t_i$ and $t_i^w$ differ among the transformers because of different dates of entry into the transformer population.

The total number of future failures between the times when the individual transformers have ages $t_i$ and $t_i^w$ is $K = \sum_{i=1}^{n^*} I_i$, where $I_i \sim \text{Bernoulli}(\rho_i)$, $i = 1, 2, \ldots, n^*$. Here $n^*$ is the number of transformers that are at risk. Thus, $K$ is a sum of independent nonidentical Bernoulli random variables. In general, there is not a simple closed-form expression for $F_K(k|\boldsymbol{\theta})$, the c.d.f. of $K$. Monte Carlo simulation can be used to evaluate the c.d.f. of $K$, to any degree of accuracy [e.g., using the algorithm in Escobar and Meeker (1999),

Section A.3]. The Monte Carlo approach, however, is computationally intensive when the number of nonidentically distributed components is large. Poisson approximation and a normal approximation based on the ordinary central limit theorem (CLT) have been suggested in the past. Here, we use an approach suggested by Volkova (1996) which is based on a refined CLT that makes a correction based on the skewness in the distribution of $K$. In particular, the estimated c.d.f. of $K$ can be approximated by

$$F_K(k|\widehat{\boldsymbol{\theta}}) = G_K\left[\frac{k+0.5-\mu_K(\widehat{\boldsymbol{\theta}})}{\sigma_K(\widehat{\boldsymbol{\theta}})}, \widehat{\boldsymbol{\theta}}\right], \qquad k = 0, 1, \ldots, n^*,$$

where $G_K(x, \widehat{\boldsymbol{\theta}}) = \Phi_{\text{nor}}(x) + \gamma_K(\widehat{\boldsymbol{\theta}})(1-x^2)\phi_{\text{nor}}(x)/6$, and

$$\mu_K(\widehat{\boldsymbol{\theta}}) = \widehat{\mathrm{E}}(K) = \sum_{i=1}^{n^*} \widehat{\rho}_i, \qquad \sigma_K(\widehat{\boldsymbol{\theta}}) = [\widehat{\mathrm{Var}}(K)]^{1/2} = \left[\sum_{i=1}^{n^*} \widehat{\rho}_i(1-\widehat{\rho}_i)\right]^{1/2},$$

$$\gamma_K(\widehat{\boldsymbol{\theta}}) = [\widehat{\mathrm{Var}}(K)]^{-3/2}\widehat{\mathrm{E}}[K-\mu_K(\widehat{\boldsymbol{\theta}})]^3 = \sigma_K^{-3}(\widehat{\boldsymbol{\theta}})\sum_{i=1}^{n^*} \widehat{\rho}_i(1-\widehat{\rho}_i)(1-2\widehat{\rho}_i)$$

are estimates of the mean, standard deviation, and skewness of the distribution of $K$, respectively.

6.2. *Calibrated prediction intervals.* The calibrated prediction interval $[\underset{\sim}{K}, \widetilde{K}]$ for the cumulative number of failures at a specified date in the future can be obtained by using the following procedure:

1. Simulate $I_i^*$ from Bernoulli$(\widehat{\rho}_i)$, $i = 1, 2, \ldots, n^*$, and compute $K^* = \sum_{i=1}^{n^*} I_i^*$.
2. Repeat step 1 $B$ times to get $K_b^*, b = 1, 2, \ldots, B$.
3. Compute $U_{Kb}^* = F_K(K_b^*|\widehat{\boldsymbol{\theta}}_b^*), b = 1, 2, \ldots, B$.
4. Let $u_K^{\mathrm{l}}, u_K^{\mathrm{u}}$ be, respectively, the lower and upper $\alpha/2$ sample quantiles of $U_{Kb}^*, b = 1, \ldots, B$. The $100(1-\alpha)\%$ calibrated prediction interval can be obtained by solving for $\underset{\sim}{K}$ and $\widetilde{K}$ in $F_K(\underset{\sim}{K}|\widehat{\boldsymbol{\theta}}) = u_K^{\mathrm{l}}$ and $F_K(\widetilde{K}|\widehat{\boldsymbol{\theta}}) = u_K^{\mathrm{u}}$, respectively.

Note that the uncertainty in $\widehat{\rho}_i$ has been accounted because $\widehat{\rho}_i$ is a function of $\widehat{\boldsymbol{\theta}}$. The uncertainty $\widehat{\boldsymbol{\theta}}$ is accounted by the bootstrap.

6.3. *Prediction results.* In this section we present the results for predicting the cumulative number of failures for the population of transformers that are at risk, based on the Weibull distribution regression model with the stratification cutting at year 1987. Figure 5 shows the predictions for the cumulative number of failures and 90% and 95% pointwise prediction intervals separately for the Old and the New groups. Note the difference in the size of the risk sets for these two groups. Figure 6 gives similar predictions for the Old and New groups combined. Figure 7 shows predictions and 90% and 95% pointwise prediction intervals for manufacturers MA (New group) and MB (Old group).

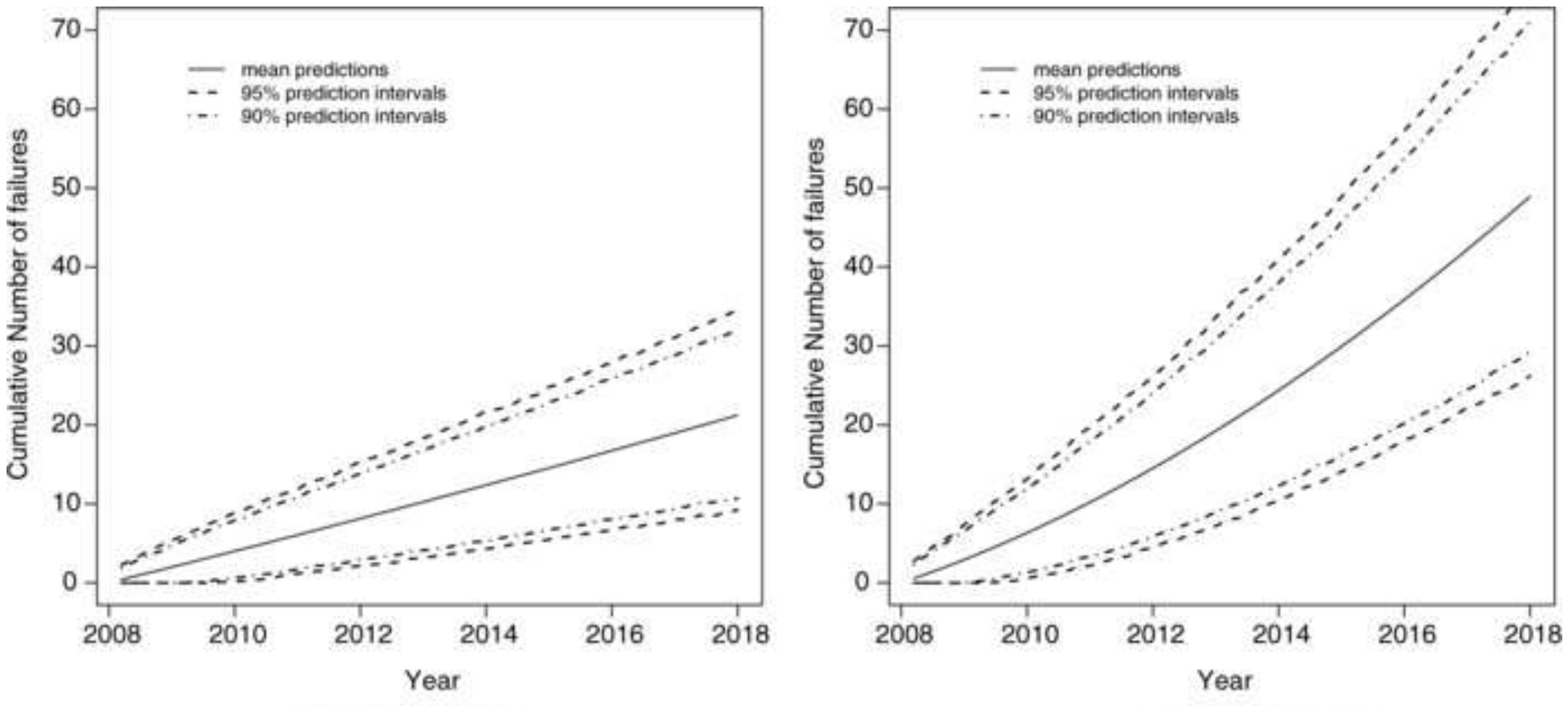

(a) Old group (b) New group

Fig. 5. *Weibull distribution predictions and prediction intervals for the cumulative number of future failures. Number of units in risk set: Old 449, New 199.*

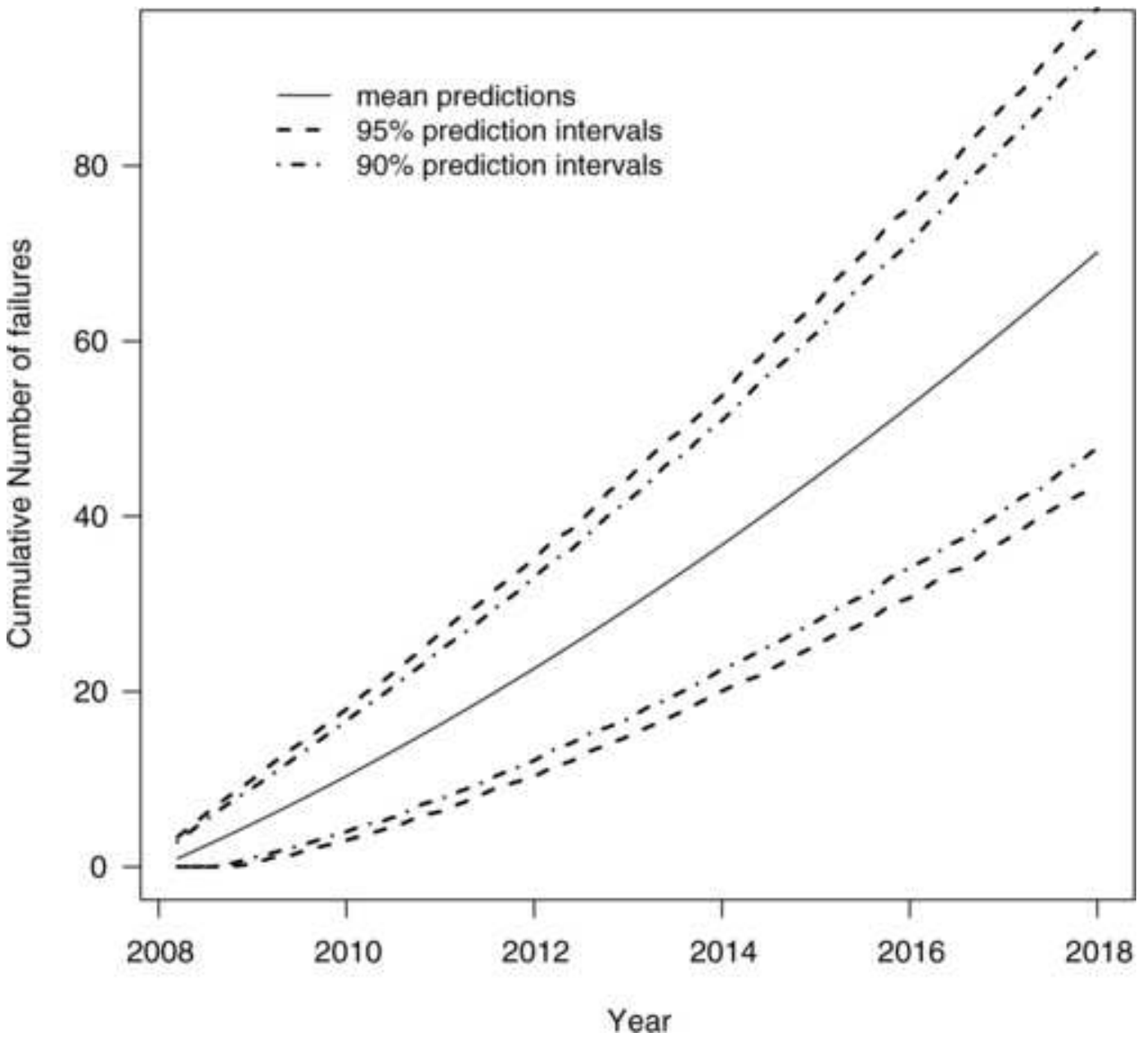

Fig. 6. *Weibull distribution predictions and prediction intervals for the cumulative number of future failures with the Old and New groups combined. 648 units in risk set.*

## 7. Sensitivity analysis and check for consistency.

7.1. *Sensitivity analysis.* The prediction interval procedures account only for statistical uncertainty. Model uncertainty (e.g., the data might be from either the Weibull or lognormal or some other distribution) is also an important source of uncertainty for the prediction. In some situations, the model uncertainty can dominate statistical uncertainty, especially when the sample size is large. Thus, when data or engineering knowledge do not unambiguously define the model, it is important to do a sensitivity analyses for the predictions by perturbing model assumptions.

*Distribution assumption.* We did sensitivity analyses to assess the effect that the assumed underlying distribution has on predictions. Figure 8 compares the predicted cumulative number of future failures and the corresponding 90% prediction intervals for the lognormal and Weibull distributions. For the Old group, the predictions are not highly sensitive to the distribution assumption. Predictions for the New group, however, are somewhat sensitive to the distribution assumption. This difference is partly due to a larger amount of extrapolation for the New group than the Old group over the next 10 years. As is generally the case with extrapolation in time, the lognormal predictions are more optimistic than the Weibull predictions.

*Cutting year.* We also did sensitivity analyses to assess the effect that using different Old/New cut points has on predictions. The results are shown in Figure 9. Changes to the cutting year have little effect in the Old group. The results in the New group are more sensitive to this choice. Note that in Figure 9b, the prediction intervals for cutting year 1990 get wider than other cutting years when time is increasing. This occurrence is caused by the fact that there is only one failure in the MC.ME.Other_New group if cutting year 1990 is used, and, thus, the random weighted bootstrap samples have more variabilities than using other cutting years. As mentioned in Section 4.1, we use 1987 as the cut year; this is on the pessimistic side of the sensitivity analysis results for the New group.

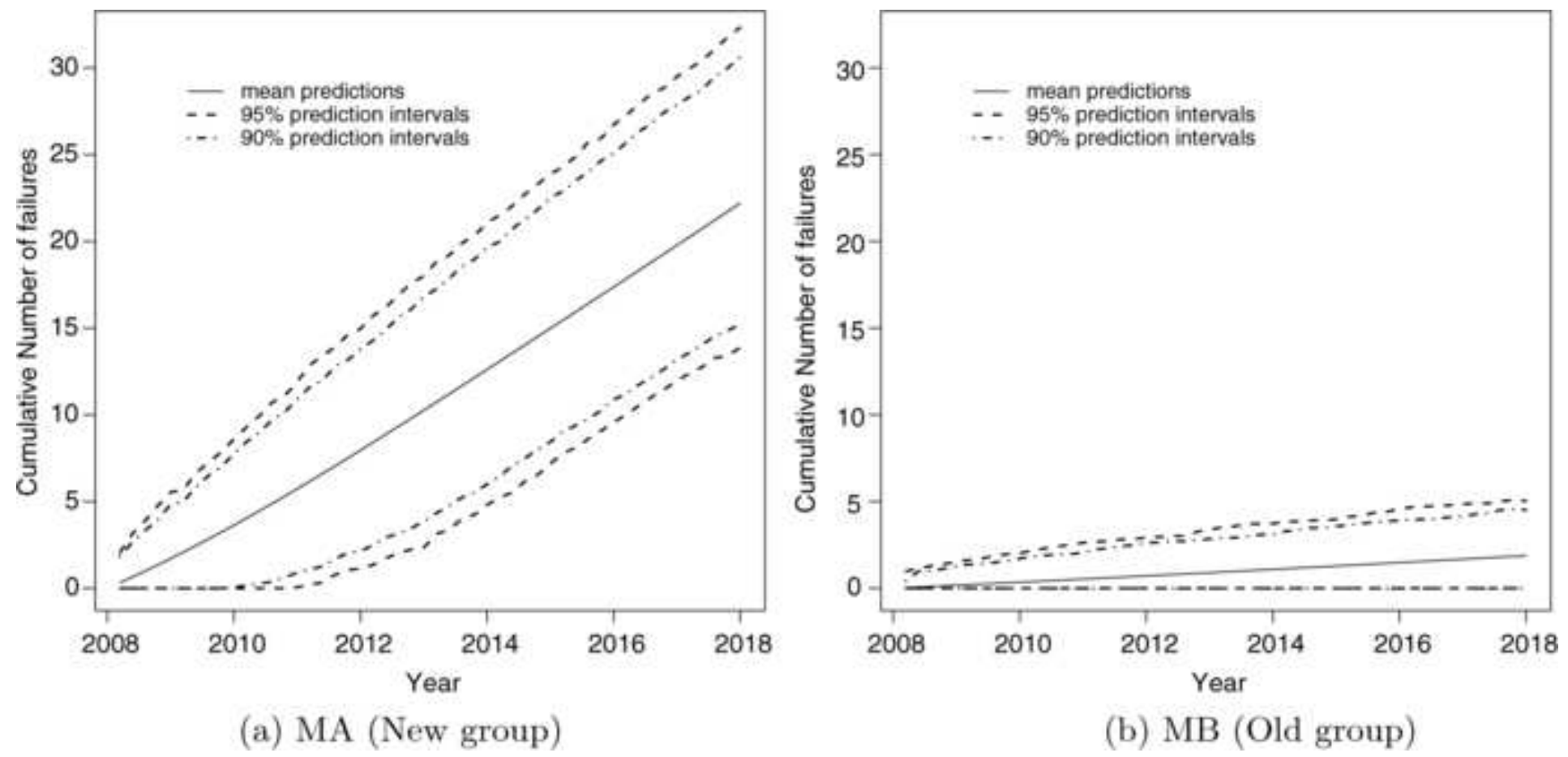

Fig. 7. *Weibull distribution predictions and prediction intervals for the cumulative number of future failures for manufacturers MA and MB. Number of units in the risk set: MA 37, MB 44.*

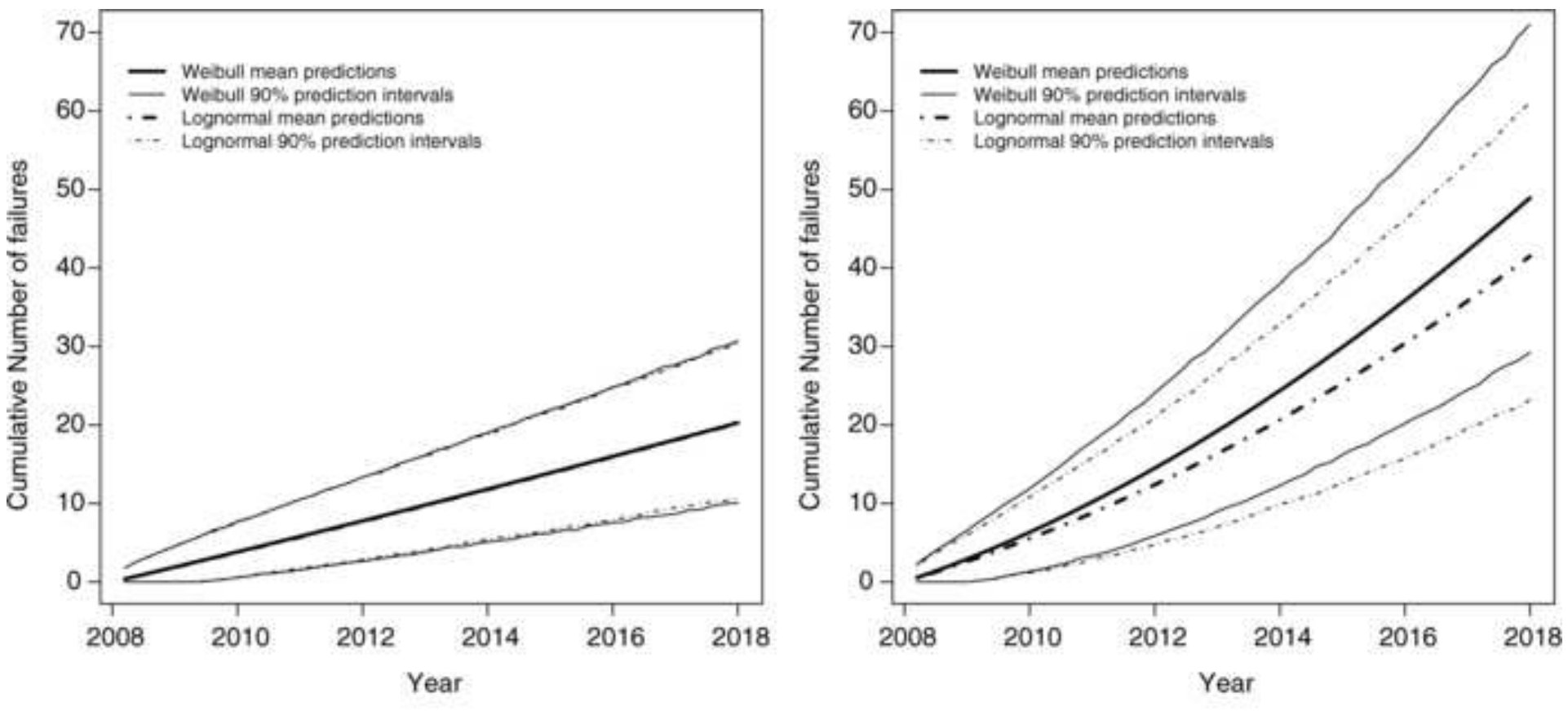

(a) Old group (b) New group

Fig. 8. *Sensitivity analysis for the effect that transformer lifetime distribution assumption has on the predicted cumulative number of future failures.*

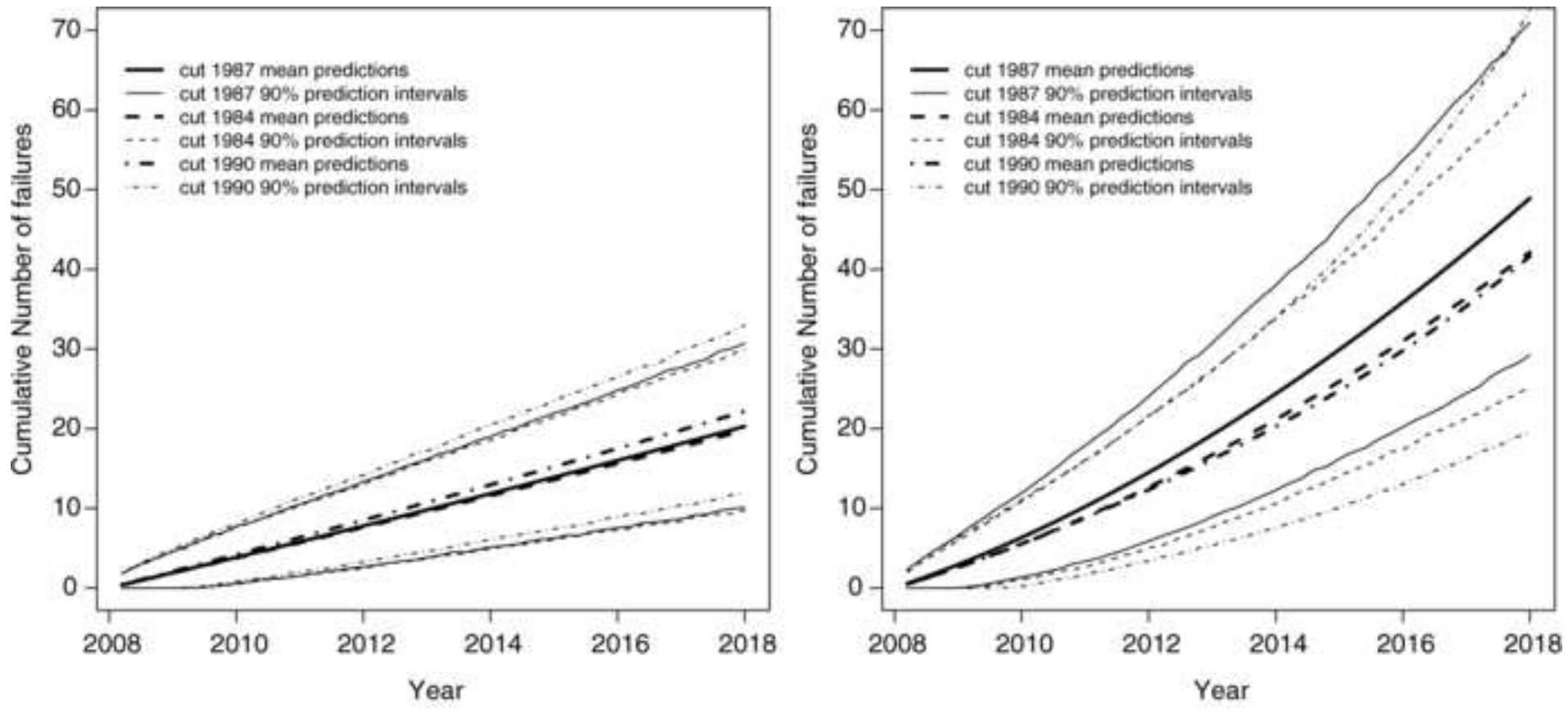

(a) Old group (b) New group

Fig. 9. *Sensitivity analysis for the effect that cutting year for the MC transformers has on the mean predicted number of failures.*

7.2. *Check for consistency.* As a part of the model diagnostics, a check for the consistency of the model was done to assess the prediction precision of the model. Generally, we would like to do this by holding out more recent failures when building the prediction model and then using the model to predict "future" failures that have already occurred. In our transformer application, however, there are not enough data to do this. Instead we used model parameter estimates based on all of the data for this check. To do the check, we move the data-freeze date back to 1994 and those units that went into service after 1994 are added into the risk set when they enter service. Then, we use our model to predict the fraction of units failing from 1994 to 2007. Figure 10 gives a plot of the predicted fraction failing and the corresponding nonparametric estimates based on the Turnbull nonparametric estimator. Figure 10 also shows 90% pointwise prediction intervals. The zigzag in the prediction intervals is caused by the new units entering into the risk set over the time period. The prediction results agree reasonably well with the nonparametric estimate. The slight disagreement in the New group (well within the prediction bounds) is due to a small difference in the behavior of the units that failed before and after the assumed 1994 data-freeze point for the check.

**8. Discussion and areas for future research.** In this paper we developed a generic statistical procedure for the reliability prediction problem that can be used with complicatedly censored and truncated data. This prediction interval procedure has broader applications, such as in field reliability prediction for warranty data [i.e., Ion et al. (2007) where only point predictions were given].

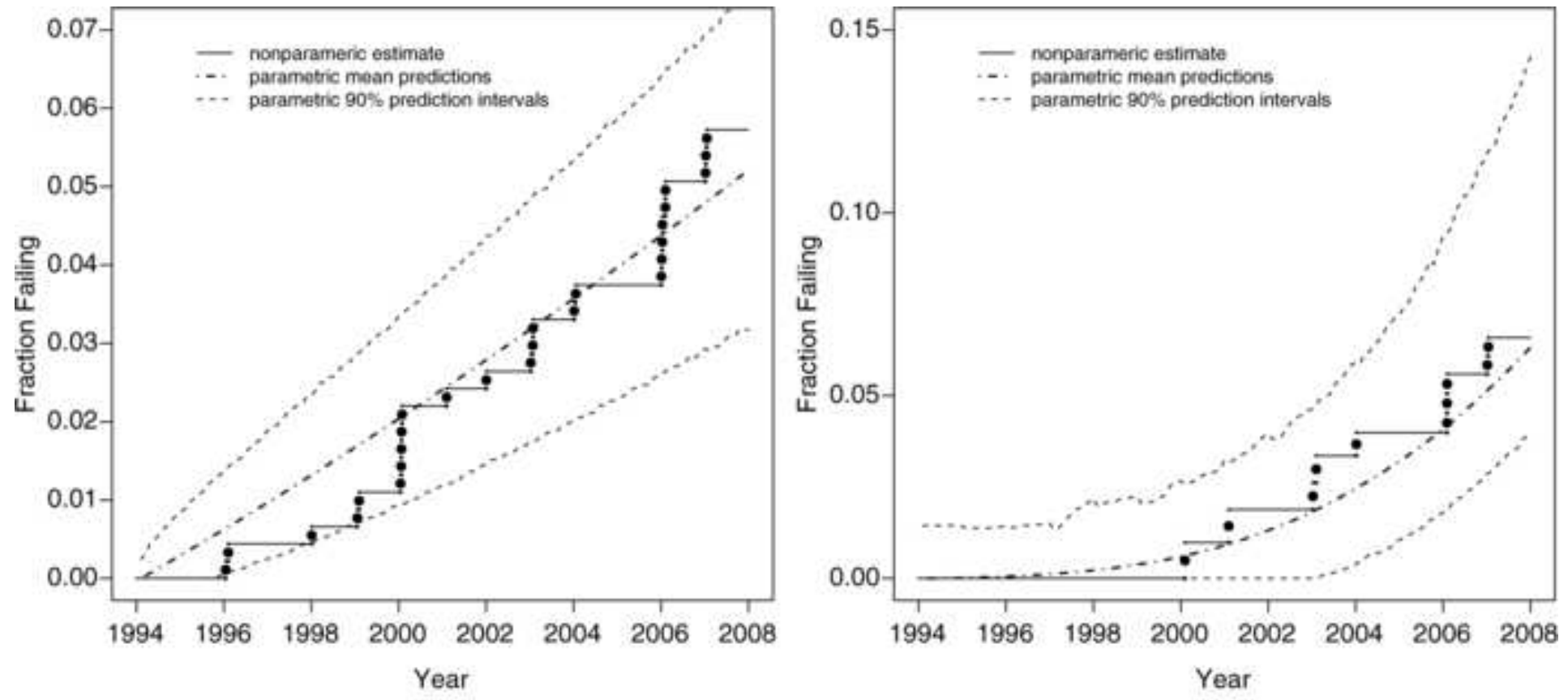

(a) Old group (b) New group

FIG. 10. *Back check of the model: parametric predictions compared with Turnbull nonparametric estimates.*

In our data analyses, we found that some transformers manufactured by particular manufacturers, for example, MA, tend to have shorter lives. We suggested that the company should pay particular attention to these transformers. Although the prediction intervals for the individual transformers are often too wide to be directly useful in determining when a transformer should be replaced, the quantitative information does provide a useful ranking for setting priorities in maintenance scheduling and for selecting transformers that need special monitoring attention or more frequent inspections to assess their health. The prediction intervals for the cumulative number of failures over time for the population of the transformers are useful for capital planning.

The prediction intervals for individual transformers tend to be wide. If usage and/or environmental information for the individual transformers were available (e.g., load, ambient temperature history, and voltage spikes, etc.), it would be possible to build a better predictive model that would more accurately predict individual lifetimes. Models in Nelson (2001) and Duchesne (2005) can be used in this direction. Further developments would, however, be needed to compute appropriate prediction interval procedures.

If engineering knowledge can provide information about the shape parameter of the lifetime distribution of the transformer or regression coefficients, the Bayesian approach could be used to take advantage of the prior information and could narrow the width of the prediction intervals. Regression analysis can be done directly on remaining life. Methods described in Chen (2007) can provide other modeling and prediction possibilities but require alternative parametric assumptions.

The transformer dataset used in our study contained only limited information about the causes of failure. As explained in Section 2.1, however, the predominant failure mechanism is related to degradation of the paper-like insulating material. In other prediction problems there can be multiple causes of failure. Particularly when these failure modes behave differently, have different costs, or when the information is to be used for engineering decisions, it is important to analyze and predict the failure modes separately (e.g., using methods similar to traditional competing risk analysis). In these applications such extension raises some interesting technical challenges, such as dealing with dependency among the failures modes that one would expect in field data. For example, it is easy to show that there will be positive dependence between failure mode lifetime distributions when analysis is done in terms of time in service when failure are driven by the amount of use and there is use-rate variability in the population.

This paper has focused on the prediction of transformer life. There are many other potential applications for this kind of work, ranging from aging aircraft to consumer products. There are also important links to the important area of *System Health Management*. In our experience, each lifetime prediction problem requires somewhat different lifetime modeling tools and methods, but the basic idea of using the distribution of remaining life for individual units in the population that are at risk for prediction is a constant.

**Acknowledgments.** We would like to thank the editor, the associate editor, and two referees, who provided comments that helped us improve this paper. We also would like to thank Luis Escobar and Katherine Meeker for helpful comments that improved this paper. The work in this paper was partially supported by funds from NSF Award CNS0540293 to Iowa State University.

## SUPPLEMENTARY MATERIAL

**Supplement to "Prediction of remaining life of power transformers based on left truncated and right censored lifetime data"** (DOI: 10.1214/08-AOAS231SUPP; .pdf). This supplement provides a description of the difficulties that we had in fitting a model to describe the failure behavior of the MD group data. The problems arise because of the large amount of truncation in this particular group.

Y. HONG
W. Q. MEEKER
DEPARTMENT OF STATISTICS
IOWA STATE UNIVERSITY
AMES, IOWA 50011
USA
E-MAIL: hong@iastate.edu
wqmeeker@iastate.edu

J. D. MCCALLEY
DEPARTMENT OF ELECTRICAL AND COMPUTER ENGINEERING
IOWA STATE UNIVERSITY
AMES, IOWA 50011
USA
E-MAIL: jdm@iastate.edu